\documentclass[12pt]{iopart}
\usepackage{iopams}  
\usepackage{graphicx,epsfig}
\usepackage{hyperref}

\newcommand{\eeq}{\end{equation}}
\newcommand{\beq}{\begin{equation}}
\newcommand{\bw}{\begin{widetext}}
\newcommand{\ew}{\end{widetext}}
\newcommand{\eea}{\end{eqnarray}}
\newcommand{\bea}{\begin{eqnarray}}
\newcommand{\ec}{\end{center}}
\newcommand{\bc}{\begin{center}}
\newcommand{\ba}{\begin{array}}
\newcommand{\ea}{\end{array}}
\newcommand{\pa}{\partial}
\newcommand{\mb}{\mathbf}

\begin{document}
\title{Renormalized cumulants and velocity derivative skewness in Kolmogorov turbulence}
\author{\normalsize Tapas Singha$^1$\thanks{Email: s.tapas@iitg.ernet.in},  Kishore Dutta$^2$\thanks{Email: kdkishore77@gmail.com}, and { Malay K. Nandy$^1$\thanks{Email: mknandy@iitg.ernet.in}}\\ \small $^1$Department of Physics, Indian Institute of Technology Guwahati, Guwahati 781 039, India\\ \small $^2$Department of Physics, Handique Girls' College, Guwahati 781 001, India} 
\vspace{10pt}

\today

\begin{abstract}
We apply a renormalized perturbative scheme on the Navier-Stokes equation for an incompressible isotropic turbulent velocity field.  This allows us to obtain the renormalized expressions for second- and third-order cumulants of the velocity derivative directly from the corresponding Feynman diagrams. The resulting expressions are integrated numerically by excluding and including the dissipation range assuming Kolmogorov and Pao's phenomenological expressions for the energy spectrum. The ensuing values for skewness are found to be $S=-0.647$ (when the dissipation range is excluded) and 
$\mathcal{S}=-0.682$ (when the dissipation is included). These estimated values are compared with various experimental, numerical, and theoretical results. 

\end{abstract}

\section{Introduction}
The turbulent flow of an incompressible fluid governed by the Navier-Stokes (NS) equation has long been considered as a challenging problem due to its inherent non-linearity and complexity \cite{Frisch,Leslie,McComb90,Lesieur}. Various important advances have been made over the last decades in understanding the statistical properties of turbulence following from the governing dynamical equation. The NS equation for an incompressible turbulent fluid is expressed as \beq \frac{\pa u_i}{\pa t}+u_j\frac{\pa u_i}{\pa x_j}=-\frac{1}{\rho}\frac{\pa p}{\pa x_i}+\nu_0\frac{\pa^2u_i}{\pa x_j\pa x_j},\label{eq:NS1} \eeq with the incompressibility condition \beq \frac{\pa u_i}{\pa x_i}=0,\eeq coming from the equation of continuity.  Here $u_i(\mb x,t)$ is the velocity field, $p(\mb x,t)$ the pressure field, $\rho$ the density, and $\nu_0$ is the kinematic viscosity of the fluid. The pressure field can be expressed in terms of the velocity field using the incompressibility condition $\partial u_i/ \partial x_i=0$. The relative importance of the inertial convective term $u_j\pa u_i/\pa x_j$ and the viscous term $\nu _0 \pa^2u_i/\pa x_j \pa x_j$ is determined by the Reynolds number $R=U L/\nu_0$, where $L$ and $U$ are the integral length and velocity scales, respectively.

In three-dimensions, the turbulent energy density obeys the Kolmogorov universal scaling (neglecting intermittency correction) \beq E(k)=C\ \varepsilon^{2/3}\ k^{-5/3}\label{Eq:energy_Kol}\eeq  in the inertial-range  $L^{-1}\ll k \ll \eta^{-1}$ where the turbulent energy cascades from the largest to the smallest scales of motion  \cite{Frisch,Leslie,Kolmogorov41}. In the above expression, $C$ is the universal Kolmogorov constant, $\varepsilon$ the energy transfer rate (per unit mass) which is also the mean dissipation rate, and $k$ is the wavenumber. The Kolmogorov microscale $\eta$, defined as $\eta=(\nu^3/\varepsilon)^{1/4}$, signifies the scale where dissipation becomes important. Within the Kolmogorov phenomenology, the eddy-viscosity follows the universal scaling \beq \nu(k)=\alpha\ \varepsilon^{1/3}\ k^{-4/3},\label{eq:nuk}\eeq where $\alpha$ is another universal constant.

Statistical characterization of NS turbulence begins most often with an equal-time $n$-th order structure function $\Phi_n(r)=\langle|\Delta \mb u_{\mb r}|^n\rangle$ that represents the $n$-th order cumulant (with respect to the probability distribution) of the velocity difference $\Delta \mb u_r=\mb u(\mb x+\mb r)-\mb u(\mb x)$ between two points separated by a displacement $\mb r$ at the same time $t$ \cite {Lesieur,She91,Kailasnath92,Giles95,Sreenivasan97,Qian98,Qian00,Li05}. Within the  Kolmogorov's phenomenological picture, the probability distribution function (PDF) in the inertial-range is universal and, as a consequence, its statistical characterization is expected to be described in terms of universal numbers \cite{Frisch,Leslie,Lesieur}. It has been observed, both  experimentally  \cite{GKBatchelor,Kuo71} and numerically \cite{She88}, that the full probability distribution deviates from the normal distribution. The increasingly non-Gaussian statistics of velocity differences towards small scales has usually been attributed to the spatially intermittent character of the fine-scale structure in such flows. As the velocity-gradient field in high Reynolds number turbulent flows is increasingly dominated by the velocity fluctuations towards the smaller scales of motion, knowledge of the statistical cumulants of velocity-gradient for a turbulent flow is important to understand the fine-scale statistics of turbulence.

In the past couple of decades, extensive experimental \cite{Frisch,Belin97,Burattini2008}, numerical \cite{Orszag72,Vincent91,Wang96,Gotoh02,Ishihara07}, and theoretical \cite{Andre77,Kraichnan90,Lesieur00,Qian94,Nelkin90,Chevillard06,YO_86,Smith92} investigations on the statistical cumulants of velocity gradient for homogeneous and isotropic turbulent flows have been carried out. Experimentally, the longitudinal velocity gradient $\pa u_1/\pa x_1$ is found to be negatively skewed \cite{Belin97,Burattini2008}, yielding the velocity gradient skewness $S=\langle(\frac{\pa u_1}{\pa x_1})^3 \rangle/\langle (\frac{\pa u_1}{\pa x_1})^2\rangle^{3/2} \approx -0.5$. This was also confirmed via a numerical  simulation with the three-dimensional NS equation for incompressible flow \cite{Orszag72} that led to $S=-0.47$ at moderate Taylor-microscale Reynolds numbers ($20\leqslant R_\lambda\leqslant 45$). Subsequently, direct numerical simulations (DNSs) for three-dimensional homogeneous isotropic turbulence \cite{Vincent91,Wang96,Gotoh02,Ishihara07} also suggested that $S$ is independent of $R_{\lambda}$ at moderate $R_{\lambda}$. At $R_\lambda\approx 150$, Vincent and Meneguzzi \cite{Vincent91} obtained $S=-0.5$ in their DNS. Wang {\it et al.}\ \cite{Wang96} performed a set of DNSs on both the freely decaying and forced stationary isotropic turbulence fields for $21<R_\lambda<195$ and showed that $S\approx-0.5$ is almost independent of the flow Reynolds number. Performing a high resolution DNS, Gotoh \cite{Gotoh02} suggested that the skewness factor of the longitudinal velocity derivative is very insensitive to $R_\lambda$ over a range $38\leqslant R_\lambda\leqslant 460$, and its average value is  $S=-0.53$. They also reported the scaling $S \propto R^{0.0370}_{\lambda}$ via least square fit of the DNS data. In another recent DNS with $4096^3$ grid points, Ishihara {\it et al.}\ \cite{Ishihara07} showed that the $S\approx -0.5$ for $R_\lambda<200$.

In theoretical investigations, namely in eddy-damped quasi-normal Markovian (EDQNM) closure \cite{Andre77,Kraichnan90,Lesieur00,Qian94}, multi-fractal (MF) model \cite{Frisch,Nelkin90,Chevillard06}, and dynamic RG analyses \cite{YO_86,Smith92}, the value of skewness turned out to be comparable to the above mentioned experimental and numerical predictions. Using EDQNM closure,  Andr{\'e} and Lesieur \cite{Andre77} showed that  the value of $S$ increases with $R_\lambda$ and tends to the value $S=-0.495$ for large $R_\lambda$. Kraichnan \cite{Kraichnan90} applied a mapping closure model on the NS equation and showed that skewness of turbulent velocity derivative is asymptotically independent of the Reynolds number.  Using the EDQNM closuer, Lesieur and Ossia \cite{Lesieur00} investigated  three-dimensional isotropic turbulence  at very high Reynolds numbers and obtained $S=-0.547$, independent of Reynolds number. Qian \cite{Qian94} used a nonequilibrium statistical mechanics closure method and obtained a constant value of skewness, namely $S=-0.515$ for very high value of Reynolds numbers. The MF model \cite{Nelkin90} suggested that the skewness increases with Reynolds number as $S\sim -R_\lambda^{0.14}$. The dynamic RG scheme of Yakhot and Orszag \cite{YO_86} yields  $S=-0.4878$ in three dimensions. Smith and Reynolds \cite{Smith92} made a correction in their calculation and obtained $S=-0.59$. These theoretical estimates for the velocity derivative skewness are comparable to the experimental estimates \cite{Frisch,Belin97,Burattini2008} and numerical predictions \cite{Vincent91,Wang96,Gotoh02}. 

There have been other theoretical attempts \cite{Tatsumi78,Kaneda93,Kida97} for the calculation of velocity derivative skewness where a relatively higher magnitude of skewness were obtained. Tatsumi {\it et al.}\ \cite{Tatsumi78}, through a multiple-scale cumulant expansion (MSCE) scheme, showed that the magnitude of skewness increases with $R_{\lambda}$ and saturates to $S=-0.65$ at very high Reynolds numbers. With a different choice for the initial energy spectrum, they obtained a slightly different value, namely, $S=-0.67$. Kaneda \cite{Kaneda93}, employing the Markovianized Lagrangian renormalized approximation (MLRA)  for freely decaying homogeneous isotropic turbulence, obtained $S=-0.66$. Kida and Goto \cite{Kida97} applied Lagrangian direct interaction approximation (LDIA) for stationary turbulence and obtained $S=-0.66$, in agreement with that of the decaying turbulence. A few recent experiments and a high-resolution DNS also suggest higher magnitudes for skewness at high Reynolds numbers. Particularly, recent hot-wire anemometer measurements in active-grid wind-tunnel turbulence \cite{Gylfason04}  yielded the velocity derivative skewness over a range $149\leqslant R_\lambda\leqslant 729$. From the measured data for $S$, they obtained an $R_\lambda$-dependent empirical relation $S=-0.33 R_\lambda^{0.09}$, indicating that the value of $S$ slowly becomes more negative with increasing Reynolds number [the skewness value is $S=-0.597$ for $R_{\lambda}=729$ (corresponding to $R \approx R_\lambda^2/16= 3.3 \times 10^4$) \cite{Lesieur}]. A similar behavior was observed in a very high resolution DNS  \cite{Ishihara07} carried out up to $R_\lambda=1130$ where the skewness data for  $200 < R_\lambda \leqslant 680$ were found to fit well with a power law $S \sim -(0.32\mp0.02) \ R_{\lambda}^{0.11\pm 0.01}$. They reported the skewness value $S=-0.648\pm0.003$ for $R_\lambda=680$ (corresponding to $R\approx 2.9\times10^4$). 

The dynamic RG scheme was initially used by Forster, Nelson and Stephen \cite{FNS} for the case of NS fluid along with the coupled problem of the advection of a passive scalar subjected to a random driving force. They adopted the procedure developed earlier by Ma and Mazenko \cite{Ma75}. It was observed by DeDominicis and Martin \cite{DeDom79} that for a particular case of randomly stirred model, Kolmogorov's inertial-range scaling for the energy spectrum, $E(k)\sim k^{-5/3} $, is realizable. Yakhot and Orszag \cite{YO_86} applied the dynamic RG scheme on the randomly stirred model of DeDominicis and Martin and calculated various universal numbers including the velocity derivative skewness associated with Kolmogorov turbulence. However, their RG estimates for the skewness was comparatively smaller in magnitude than that of the estimates coming from  MSCE \cite{Tatsumi78},  LRA \cite{Kaneda93} and LDIA \cite{Kida97}, and the estimates from high resolution DNS \cite{Ishihara07}.  Here we consider an alternative scheme that yields renormalized quantities relevant for the calculation of skewness. This scheme is different from the above RG schemes as it finds a relation between the renormalized Feynman diagrams involving the renormalized viscosity. Recently this procedure was found to be successful in determining the experimentally observed statistical characteristics in Kardar-Parisi-Zhang (KPZ) \cite{Singha-Nandy14_PRE.90.062402,Singha-Nandy15_jstatmech,Singha-Nandy_KPZ2D_JSTAT} and Villain-Lai-Das Sarma (VLDS) \cite{Singha-Nandy16_jstatmech} surface growth dynamics. This scheme enables us to calculate the second-and third-order cumulants of velocity derivative and the resulting value for skewness is obtained as $S=-0.647$. This value is obtained when only the inertial range with Kolmogorov scaling is considered. It is however important to take the dissipation range into account in order to calculate the integrals for the second- and third-order cumulants. We employ Pao's model \cite{Pao_POF_1965} that joins the inertial range smoothly with the dissipation range and evaluate the integrals for the second- and third-order cumulants. The resulting skewness value $\mathcal{S}$ turns out to be $\mathcal{S}=-0.682$. These estimates are closely comparable to other theoretical estimates coming from MSCE \cite{Tatsumi78}, LRA \cite{Kaneda93}, and LDIA \cite{Kida97} as well as the  estimates from a recent high resolution DNS  \cite{Ishihara07}.

The paper is organized as follows. In Section\@ II we introduce the randomly stirred model and calculate an amplitude ratio needed later. Calculations of second- and third-order renormalized cumulants and skewness in the Kolmogorov range are presented in Section\@ III. Section\@ IV generalizes the calculations of second- and third-order cumulants and skewness to include the dissipation range. Finally, a discussion and conclusion are given in Section\@ V. All the technical details involving intermediate steps of the calculations are given in the appendix.  
\section{Randomly stirred dynamics}\label{sec:RSD}
In order to calculate the velocity derivative skewness, we use Fourier transformation of the velocity field \beq u_i(\mb x,t)= \int\frac{d^d k}{[2\pi]^d}\int_{-\infty}^{\infty}\frac{d\omega} {[2\pi]}u_i(\mb k,\omega) e^{i(\mb k\cdot\mb x-\omega t)}, \label{eq:Fourier} \eeq  along with the incompressibility condition \[ k_i u_i(\mb k,\omega)=0.\] Thus, the Fourier transformed NS equation becomes \beq \left(-i\omega+\nu_0 k^2\right)u_i(\mb k,\omega)=f_i(\mb k,\omega)-\frac{i\lambda_0}{2}P_{ijl}(\mb k) \displaystyle\int\frac{d^d q}{[2\pi]^d}\int_{-\infty}^{\infty}\frac{d\Omega}{[2\pi]}u_j(\mb q,\Omega)u_l(\mb k-\mb q,\omega-\Omega), \label{eq:NSk}\eeq where $P_{ijl}(\mb k)=k_jP_{il}({\bf k})+k_lP_{ij}({\bf k})$ 
with $P_{ij}({\bf k})=\delta_{ij}-k_ik_j/\bf k^2$. A random force term $f_i(\mb k,\omega)$ is introduced 
in Eq.\ (\ref{eq:NSk}) following the randomly stirred model of DeDominicis and Martin \cite{DeDom79}. This forcing field maintains a statistically steady state and it is assumed to have a Gaussian white-noise statistics with the correlation \beq \langle f_i(\mb k,\omega)f_j(\mb k',\omega')\rangle = F(k)P_{ij}(\mb k)[2\pi]^d \delta^d(\mb k+\mb k')[2\pi]\delta(\omega+\omega'), \label{eq:force}\eeq where $F(k)$ is modeled as \beq F(k)=\frac{2D_0}{k^{y}}, \label{eq:noise} \eeq with $d$ the space dimension, $D_0$ a constant and $y$ is a parameter which is taken as $y=d$ for consistency with Kolmogorov spectrum given by Eq.\ (\ref{Eq:energy_Kol}). An expansion parameter $\lambda_0(=1)$  is introduced in the non-linear term of Eq.\ (\ref{eq:NSk}). 
\begin{figure}[!]
\centering 
\includegraphics[width=0.4\textwidth]{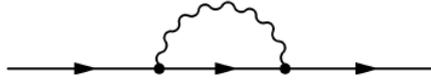} \caption{Feynman diagram giving contribution to $\Upsilon$. The propagators are indicated by solid lines and correlation by a wiggly line.}
\label{fig:alphaC} 
\end{figure}

Since the nonlinear term poses mathematical difficulty in the problem, it is customary  to treat it as a perturbation. It is in fact possible to construct renormalized expressions for the second and third cumulants from the perturbation theory.  We shall calculate the renormalized cumulants in the next two sections. We shall see that the expressions for the second and third cumulants contain the amplitude ratio $\alpha^2/C$. This amplitude ratio can be obtained via a recursive shell elimination procedure as follows. The value of $\nu(k)k^2$, where $\nu(k)$ is the renormalized viscosity, is determined by the renormalized loop in Fig.\ \ref{fig:alphaC}. The bare value of the loop is determined by the expression \beq \Upsilon=2\lambda_0^2 D_0 P_{ijl}(\mb k) \int \frac{d^{d+1} \hat q}{[2\pi]^{d+1}} P_{jni}(\mb k-\mb q) P_{ln}(\mb q) |\mb q|^{-y}|G_0(\hat{q})|^2 G_0(\hat{k}-\hat{q}),\eeq where we use the short-hand notation $\displaystyle \int d^{d+1} \hat q\equiv \int d^d q \int^{\infty}_{-\infty}  d\Omega$ and $G_0(\hat q)\equiv G_0(\mb q,\Omega)=[-i\Omega+\nu_0q^2]^{-1}$ represents the bare propagator. We perform the integration over frequency $\Omega$ and expand the integrand in the limit $q \gg k$ because we wish to eliminate the high wavenumber band $\Lambda_0 e^{-r} \leqslant q \leqslant \Lambda_0$, where $\Lambda_0$ is the ultraviolet cut-off. Consequently, we find that $\Upsilon$ is proportional to $k^2$ in the large-scale long-time limit. The quantity $\nu^{<}(r)= \Upsilon^{<}(r)/k^2$, where $\Upsilon^{<}(r)$ is the value of $\Upsilon$ after the elimination of the high wavenumber band, is given by \beq \nu^{<}(r)=\frac{S_d}{[2\pi]^{d}}\left(\frac{d^2-4-y}{2d(d+2)}\right) \left(\frac{\lambda^2_0 D_0}{\nu^2_0}\right) \int^{\Lambda_0}_{\Lambda_0 e^{-r}} q^{d-y-5}dq, \eeq where $S_d=2\pi^{d/2}/\Gamma(d/2)$ is the surface area of an unit sphere embedded in $d$-dimensional space. Assuming that the wave number band is eliminated in recursive steps, we find a differential equation from the above expression as \beq \frac{d \nu}{dr}= \frac{S_d}{[2\pi]^{d}}  \left[\frac{d^2-4-y}{2d(d+2)} \right] \left[\frac{\lambda^2_0 D_0}{\nu^2(r)\Lambda^{4+y-d}(r)}\right], \eeq where we write $\Lambda_0 e^{-r}=\Lambda(r)$. This yields \beq \nu^3(r)=\frac{S_d}{[2\pi]^d}\left[\frac{d^2-4-y}{2d(d+2)}\right]\left(\frac{3\lambda^2_0 D_0}{4+y-d}\right)\Lambda^{(d-y-4)}(r).\eeq
Using Eq.\ (\ref{eq:nuk}) and with the identification $k=\Lambda(r)$, we see that the consistency in scaling is obtained for $y=d$. Thus setting $\lambda_0=1$, we obtain \beq \alpha^3 \varepsilon=\frac{S_d}{[2\pi]^d}\left[\frac{d^2-4-y}{2d(d+2)}\right]\left(\frac{3}{4+y-d}\right)D_0. \eeq  From the scaling relations given by Eqs.\ (\ref{Eq:energy_Kol}) and (\ref{eq:nuk}), the noise amplitude $D_0$ can be obtained as $D_0=2 \pi^2 \alpha C \varepsilon$ for $d=3$. We thus obtain \beq \frac{\alpha^2}{C}=0.050\label{alpha2C}\eeq for $d=3$. This numerical value is consistent with the EDQNM prediction, namely, $\alpha=0.28$ for $C=1.6$ \cite{Chollet81}, as indicated in Ref. \cite{Smith92}. 

\section{Evaluation of skewness in the Kolmogorov range}
In this section, we shall provide the essential steps involved in the calculation of the velocity derivative skewness from the renormalized perturbative scheme, the technical details of the calculations are given in the appendix. Here we assume that the Kolmogorov scaling given by Eq.\ (\ref{Eq:energy_Kol}) is valid in the inertial-range and  neglect the small correction (to the $-5/3$ exponent) due to intermittency.

\subsection{The Second Cumulant of Velocity Derivative}

\begin{figure}[!]
\centering 
\includegraphics[width=0.5\textwidth]{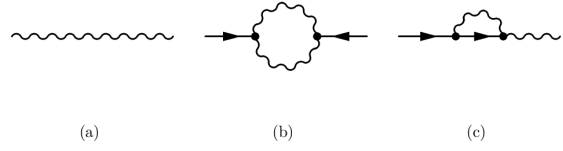} \caption{Feynman diagrams for the second cumulant $W_2$. Fig.(a) and (b) correspond to $W_2^{(1)}$ and $W_2^{(2)}$ respectively.} \label{fig:w2} 
\end{figure}

The second cumulant of the derivative of fluctuating velocity distribution is defined as \beq W_2=\left \langle \left(\frac{\pa u_1(\mb x,t)}{\pa x_1}\right)^2 \right \rangle -\left \langle \frac{\pa u_1(\mb x,t)}{\pa x_1 } \right \rangle^2.
\label{eq:W20} \eeq With the assumption of homogeneity and isotropy, the ensemble average $\langle u_{i}(\mb x,t) \rangle=0$, so that $\langle \pa u_1(\mb x,t)/ \pa x_1\rangle=0$. Thus we write Eq.\ (\ref{eq:W20}) as 
\beq W_2=\left \langle \left(\frac{\pa u_1(\mb x,t)}{\pa x_1}\right)^2 \right \rangle=\left \langle \left(\frac{\pa u_3(\mb x,t)}{\pa x_3}\right)^2 \right \rangle=W_2^{(1)}+W_2^{(2)}, \eeq where the second equality follows from the assumption of homogeneity and isotropy. $W_2^{(1)}$ and $W_2^{(2)}$ are contributions coming from $O[\lambda_0^{(0)}]$ and $O[\lambda_0^{(2)}]$ terms of the perturbation series due to the elimination of velocity fluctuations belonging to the shell $\Lambda_0e^{-r}\leqslant q\leqslant\Lambda_0$. Figs.\ \ref{fig:w2}(a) and \ref{fig:w2}(b) represent the contributions $W_2^{(1)}$ and $W_2^{(2)}$, respectively. The corresponding expressions are written as 
\beq W_2^{(1)}= -\int \frac{d^{d+1} \hat{k}}{[2\pi]^{d+1}} \int \frac{d^{d+1} \hat{k}'}{[2\pi]^{d+1}} k_3 \ k'_3 \ \langle u_3(\mb k,\omega) u_3(\mb k',\omega')\rangle, \label{W21-Fourier}\eeq and \bea W^{(2)}_2= && \left(\frac{\lambda_0}{2}\right)^2 \int \frac{d^{d+1} \hat{k}}{[2\pi]^{d+1}} \int \frac{d^{d+1} \hat{k}'}{[2\pi]^{d+1}} k_3 \, k'_3 \, P_{3mn}(\mb k) P_{3sl}(\mb k') G(\hat{k})G(\hat{k'})  \nonumber \\ && \int \frac{d^{d+1} \hat{p}}{[2\pi]^{d+1}} \int \frac{d^{d+1} \hat{q}}{[2\pi]^{d+1}} \langle u_m(\hat p)u_n(\hat k-\hat p) u_s(\hat q) u_l(\hat k'-\hat q) \rangle. \label{eq:W22-Fourier} \eea Assuming that the flow field is statistically homogeneous in space and stationary in time, we can express the velocity correlation in terms of the renormalized quantities as \beq \langle u_i(\mb k,\omega) u_j(\mb k',\omega')\rangle = 2D_0 |\mb k|^{-y} P_{ij}(\mb k) |G(\mb k,\omega)|^2 [2\pi]^{d+1}\delta^d(\mb k+\mb k')\delta(\omega+\omega').  \label{vel-corre} \eeq Thus we obtain Eq.\ (\ref{W21-Fourier}) as  \beq W_2^{(1)}=2 D_0\int \frac{d^{d+1} \hat{k}}{[2\pi]^{d+1}} k^{-y} P_{33}(\mb k)\ k_3^2 \ |G(k,\omega)|^2, \label{eq:W21-Fourier1}\eeq  and Eq.\ (\ref{eq:W22-Fourier}) as \beq W_2^{(2)}=\int \frac{d^{d+1} \hat{k}}{[2\pi]^{d+1}} k^2_3 P_{3mn}(\mb k) P_{3sl}(\mb k) |G(k,\omega)|^2 K_{mnsl}^{(2)}(\mb k,\omega) \label{eq:W22}.\eeq The un-renormalized expression for $K_{mnsl}^{(2)}(\mb k,\omega)$ is given by  \beq K_{mnsl}^{(2)}(\mb k,\omega)=2 \lambda_0^2 D_0^2 \int \frac{d^d p}{[2\pi]^{d}} \int  \frac{d \Omega}{[2\pi]} |\mb p|^{-2 y} P_{ms}(\mathbf{p}) P_{nl}(\mathbf{p}) |G(p,\Omega)|^4 \label{eq:L2}\eeq which represents the loop diagram without the external legs in Fig.\ \ref{fig:w2}(b).

Performing  angular and wave vector integrations in Eq.\ (\ref{eq:W21-Fourier1})  and substituting $D_0=2\pi^2\alpha C \varepsilon$, we arrive at (see Appendix A) \beq W_2^{(1)}= \frac{1}{10} C \varepsilon^{2/3} \Lambda_0^{4/3} \label{eq:w21_3}\eeq in three dimensions. In order to evaluate $W_2^{(2)}$, we first derive a differential equation for $K^{(2)<}_{mnsl}(r)$ from Eq.\ (\ref{eq:L2}) and then use dynamic scaling to construct $k$ and $\omega$ dependences of $K^{(2)}_{mnsl}(k, \omega)$ (the details are given in Appendix B). This finally yields \beq W^{(2)}_2 = \frac{7}{3000 (\alpha^2/C)}\ C  \varepsilon^{2/3} \Lambda^{4/3}_0 \label{eq:w22_3} \eeq in three dimensions. Thus, adding the contributions $W_2^{(1)}$ and $W_2^{(2)}$ from Eqs.\ (\ref{eq:w21_3}) and (\ref{eq:w22_3}), we obtain the second cumulant of velocity derivative as \beq W_2= \left[1+\frac{7}{300 (\alpha^2/C)}\right] \ \frac{C}{10} \varepsilon^{2/3} \Lambda^{4/3}_0. \label{eq:w2_final} \eeq 
\subsection{The Third Cumulant of Velocity Derivative}
\begin{figure}[!]
\centering
\includegraphics[width=0.8\textwidth]{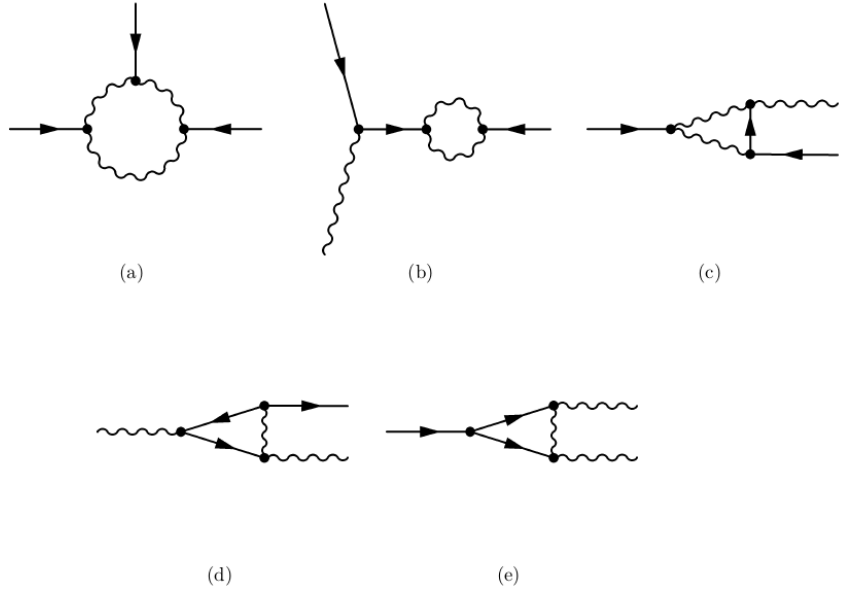}
\caption{Feynman diagrams for the third cumulant $W_3$. Fig.\ (a), (b), (d), and (e) correspond to non-zero contributions, namely, $W_3^{(1)}$, $W_3^{(2)}$, $W_3^{(3)}$, and $W_3^{(4)}$, respectively. Fig. (c) gives vanishing contribution.} \label{fig:w3} 
\end{figure}

The third cumulant of the velocity derivative $W_3=\langle\left(\frac{\pa u_1}{\pa x_1}\right)^3\rangle$ can be expressed as \beq W_3=\left\langle\left(\frac{\pa u_3}{\pa x_3}\right)^3\right\rangle= W_3^{(1)}+W_3^{(2)}+W_3^{(3)}+W_3^{(4)},\eeq where $W_3^{(1)}$, $W_3^{(2)}$, $W_3^{(3)}$, and $W_3^{(4)}$ are $[O(\lambda_0^3)]$ non-zero contributions coming from the perturbation series. The Feynman diagrams given in Figs.\ \ref{fig:w3}(a), \ref{fig:w3}(b), \ref{fig:w3}(d), and \ref{fig:w3}(e) correspond to $W_3^{(1)}$, $W_3^{(2)}$, $W_3^{(3)}$, and $W_3^{(4)}$, respectively. The other diagram, namely, Fig.\ \ref{fig:w3}(c) gives vanishing contribution in the large-scale limit. Here, we evaluate separately the contributions coming from $W_3^{(1)}$, $W_3^{(2)}$, $W_3^{(3)}$, and $W_3^{(4)}$. The detail calculations are presented in the Appendix C-F.  We shall see that  Figs.\ \ref{fig:w3}(d) and \ref{fig:w3}(e), corresponding to $W_3^{(3)}$ and $W_3^{(4)}$,  yield logarithmic contributions of opposite signs and equal magnitude and thus they cancel each other out. Consequently, the contribution to $W_3$ comes only from the two diagrams, namely, Figs.\ \ref{fig:w3}(a) and \ref{fig:w3}(b). 

The integral expression for $W_3^{(1)}$ in the Fourier space is given by \beq W_3^{(1)}=-i\int \frac{d^{d+1} \hat k}{[2\pi]^{d+1}} \int\frac{d^{d+1}\hat k'}{[2\pi]^{d+1}}  \ k_3 \ k'_3(-k_3-k_3')\ \langle  u_3(\hat k)  u_3(\hat k') u_3(-\hat k-\hat k') \rangle. \label{eq:w3}\eeq The three-point velocity correlation appearing here can be expressed in terms of the renormalized quantities as \[ \langle u_3(\hat k) u_3(\hat k') u_3(\hat k'') \rangle = P_{3mn}(\mb k) P_{3ij}(\mb k') P_{3sl}(\mb k'') L^{(1)}_{ijlmns}(\hat k,\hat k')\] \beq G(\hat k) G(\hat k') G(\hat k'') (2\pi)^{d+1} \delta^{d+1}(\hat k+\hat k'+\hat k''),\label{eq:vvv}\eeq 
where $L^{(1)}_{ijlmns}$ represents the renormalized amputated part of the loop diagram. Substituting Eq.\ (\ref{eq:vvv}) in Eq.\ (\ref{eq:w3}), we obtain \[W_3^{(1)}=-i \int \frac{d^d\mb k d\omega}{[2\pi]^{d+1}} \int\frac{d^d\mb k'd\omega'}{[2\pi]^{d+1}} k_3 \ k'_3 (-k_3-k_3') P_{3mn}(\mb k) P_{3ij}(\mb k') P_{3sl}(-\mb k-\mb k')\] \beq 
G(\mb k,\omega)\ G(\mb k',\omega') L^{(1)}_{ijlmns}(\mb k,\omega,\mb k',\omega')\ G(-\mb k-\mb k',-\omega-\omega'). 
\label{eq:w3a} \eeq The bare value of $L^{(1)}_{ijlmns}(\mb k,\omega,\mb k',\omega')$ is given by  \[ L^{(1)}_{ijlmns}(\mb k,\omega,\mb k',\omega')= 8 \left(\frac{-i\lambda_0}{2}\right)^3 \int \frac{d^{d+1}\hat q}{[2\pi]^{d+1}} \int\frac{d^{d+1}\hat q'}{(2\pi)^{d+1}} \int\frac{d^{d+1}\hat Q}{[2\pi]^{d+1}}  G_0(\hat{q})G_0(\hat{k}-\hat{q}) G_0(\hat{q'})\] \beq G_0(\hat{k'}-\hat{q'}) G_0(\hat{Q}) G_0(-\hat{k}-\hat{k'}-\hat{Q}) \langle f_m(\hat{q})f_j(\hat{k'}-\hat{q'})\rangle \langle f_n(\hat{k}-\hat{q})f_s(\hat{Q})\rangle \langle f_i(\hat{q'})f_l(-\hat{k}-\hat{k'}-\hat{Q})\rangle. \label{eq:L0}\eeq
Performing  angular and wave vector integrations and using dynamic scaling (the details are given in the appendix), we arrive at \beq W_3^{(1)}= \frac{51c_1}{15680 \pi^2} \ \frac{\varepsilon \ C^{3/2} \Lambda^2_0}{(\alpha^2/C)^{3/2}}, \label{eq:final_W31} \eeq where $c_1$ is a constant evaluated via numerical integration (see Appendix C).    

The integral expression for $W_3^{(2)}$ is given by  \[ W_3^{(2)}= \left(\frac{\lambda_0}{2}\right)^3 \int \frac{d^{d+1}\hat{k}}{[2\pi]^{d+1}} \int \frac{d^{d+1} \hat{k'}}{[2\pi]^{d+1}}\int \frac{d^{d+1}\hat{k''}}{[2\pi]^{d+1}} k_3 \ k'_3 \ k''_3 G(\hat{k'}) G(\hat{k''}) P_{3mn}(\mb k') P_{3ij}(\mb k'') \]\beq\int \frac{d^{d+1} \hat{Q}}{[2\pi]^{d+1}} \int \frac{d^{d+1} \hat{q}}{[2\pi]^{d+1}} \int \frac{d^{d+1} \hat q}{[2\pi]^{d+1}} G(\hat{Q}) P_{iab}(\mb Q) \langle u_3(\hat{k}) u_m(\hat{q}) u_n(\hat{k'}-\hat{q}) u_j(\hat{k''}-\hat{Q}) u_a(\hat{q}) u_b(\hat{Q}-\hat{q})\rangle. \label{eq:W3^2}\eeq The velocity correlation appearing in Eq.\ (\ref{eq:W3^2}) is expressed as\[ \langle u_3(\hat{k}) u_m(\hat{q}) u_n(\hat{k'}-\hat{q}) u_j(\hat{k''}-\hat{Q}) u_a(\hat{q}) u_b(\hat{Q}-\hat{q}) \rangle =24 (2D_0)^3\ (2\pi)^{3(d+1)}\delta^{d+1}(\hat k+\hat k'+\hat k'') P_{ma}(\mb q) \]\beq P_{nb}(\mb k'-\mb q)  P_{3j}(\mb k) |G(\hat q)|^2 |G(\hat k'-\hat q)|^2 |G(\hat k)|^2\ |\mb q|^{-y} |\mb k'-\mb q|^{-y}  |\mb k|^{-y}, \eeq giving \[W_3^{(2)}=24 (\lambda_0 D_0)^3 \int \frac{d^{d+1}\hat{k}}{[2\pi]^{d+1}} \int \frac{d^{d+1} \hat{p}}{[2\pi]^{d+1}} k_3 p_3 (k_3+p_3)P_{3mn}(\mb p)P_{3ij}(\mb k+\mb p)\]\beq  P_{iab}(-\mb p)] P_{3j}(\mb k) |\mb k|^{-y}  [|G(\hat p)|^2 |G(\hat k)|^2 G(-\hat k-\hat p) L^{(2)}_{mnab}(\mb p,\Omega), \label{eq:w3b}\eeq where $L^{(2)}_{mnab}(\mb p,\Omega)$ comes from the amputated part of the loop diagram in Fig.\ \ref{fig:w3}(b). The bare value of $L^{(2)}_{mnab}(\mb p,\Omega)$ is given by \beq L^{(2)}_{mnab}(\mb p,\Omega)=\int \frac{d^d q}{[2\pi]^d}\int \frac{d\Omega}{[2\pi]} |G(q,\Omega)|^2|G(p-q,\omega-\Omega)|^2 P_{ma}(\mb q) P_{nb}(\mb p-\mb q) |\mb q|^{-y}|\mb p-\mb q|^{-y}. \label{eq:calL}\eeq Using a similar procedure as before, we finally arrive at \beq W_3^{(2)}=\frac{7c_2}{200 \pi^2} \frac{\varepsilon \  C^{3/2}}{(\alpha^2/C)^{3/2}} \, \Lambda^2_0, \label{eq:final_W32} \eeq where $c_2$ is coming from numerical integration (see Appendix D).

The contribution $W_3^{(3)}$ comes from the one-loop Feynman diagram shown in Fig.\ \ref{fig:w3}(d). The corresponding integral expression can be written as \[W^{(3)}_3 =48 \left(\frac{\lambda_0}{2}\right)^3 (2D_0)^2 \int \frac{d^{d+1} \hat{k}}{[2\pi]^{d+1}} \int \frac{d^{d+1} \hat{k'}}{[2\pi]^{d+1}}[k_3 k'_3 (-k_3-k'_3)] G(-\hat{k}-\hat{k'}) P_{3ij}(-\mb k-\mb k')  \]\beq |G(\hat{k'})|^2 |G(\hat{k})|^2 [P_{3b}(\mb k') P_{3n}(\mb k)] |k|^{-y} |k'|^{-y } L^{(3)}_{ibnj}(\hat{k'}), \label{eq:W33} \eeq where the amputated part $L^{(3)}_{ibnj}(\hat{k'})$ is given by \beq L^{(3)}_{ibnj}(\hat{k'})=2 D_0 \int \frac{d^{d+1} \hat{q}}{[2\pi]^{d+1}} |\mb q+\mb k|^{-y} G(\hat{q}) G(\hat{q}-\hat{k'})|G(\hat{q}+\hat{k})|^2 [P_{iab}(\mb q-\mb k') P_{amn}(\mb q) P_{mj}(\mb q+\mb k)].  \label{eq:L3k'} \eeq As shown in the Appendix E, this yields logarithmic contribution \[ W^{(3)}_3 =-\frac{12}{d(d+2)} \frac{S_d}{[2\pi]^d}  \left(2 \pi^2 C \varepsilon^{2/3}\right)^3 \int \frac{d^{d+1} \hat{k}}{[2\pi]^{d+1}} \int \frac{d^{d+1} \hat{k'}}{[2\pi]^{d+1}} [k_3 k'_3 (k_3+k'_3)] G(-\hat{k}-\hat{k'}) P_{3ij}(\mb k+\mb k')  \]\beq |G(\hat{k'})|^2 |G(\hat{k})|^2 [P_{3i}(\mb k') P_{3j}(\mb k)] \, |\mb k|^{-y} |\mb k'|^{-y } \, \ln\left(\Lambda_0/k'\right).\label{eq:w3_3F} \eeq

The contribution $W_3^{(4)}$ comes from the one-loop Feynman diagram shown in Fig.\ \ref{fig:w3}(e). The corresponding integral expression is written as  \[ W^{(4)}_3 =24 \left(\frac{\lambda_0}{2}\right)^3 (2D_0)^2 \int \frac{d^{d+1} \hat{k}}{[2\pi]^{d+1}} \int \frac{d^{d+1} \hat{k'}}{[2\pi]^{d+1}}[k_3 k'_3 (-k_3-k'_3)] G(-\hat{k}-\hat{k'}) P_{3ij}(-\mb k-\mb k')\]\beq |G(\hat{k'})|^2 |G(\hat{k})|^2 [P_{3b}(\mb k) P_{3n}(\mb k')] |\mb k|^{-y} |\mb k'|^{-y } L^{(4)}_{ibnj}(\hat{k},\hat{k'})\label{eq:W34} \eeq where the amputated part of the loop diagram is given by \beq L^{(4)}_{ibnj}(\hat{k},\hat{k'}) = 2D_0 \int \frac{d^{d+1} \hat{Q}}{[2\pi]^d} G(\hat{Q}) G(-\hat{k}-\hat{k'}-\hat{Q}) |G(\hat{k}+\hat{Q})|^2 P_{jmn}(-\mb k-\mb k'-\mb Q) P_{iab}(\mb Q) P_{ma}(\mb k+\mb Q). \label{eq:L4kk'}\eeq  This gives (see Appendix F) \[ W^{(4)}_3=  \frac{12}{d(d+2)} \frac{S_d}{[2\pi]^d} \left(2 \pi^2 C \varepsilon^{2/3}\right)^3 \int \frac{d^{d+1} \hat{k}}{[2\pi]^{d+1}} \int \frac{d^{d+1} \hat{k'}}{[2\pi]^{d+1}} [k_3 k'_3 (k_3+k'_3)] G(-\hat{k}-\hat{k'}) P_{3ij}(\mb k+\mb k') \]\beq |G(\hat{k'})|^2 |G(\hat{k})|^2  [P_{3i}(\mb k') P_{3j}(\mb k)] \, |\mb k|^{-y} |\mb k'|^{-y } \, \ln\left(\Lambda_0/k'\right), \label{eq:w3_4F}\eeq which is also a logarithmic contribution to the third cumulant of the velocity derivative. Thus, we see the logarithmic contributions $W^{(3)}_3$ and $W^{(4)}_3$, given by Eqs.\ (\ref{eq:w3_3F}) and (\ref{eq:w3_4F}), respectively cancel each other out. 

\subsection{Inertial range skewness}
Adding the two non-vanishing contributions, namely, $W_3^{(1)}$ and $W_3^{(2)}$ from Eqs.\ (\ref{eq:final_W31}) and (\ref{eq:final_W32}), we obtain the third cumulant for the fluctuating velocity derivative as \beq W_3= \frac{1}{40 \pi^2} \left(\frac{51 c_1}{392} +\frac{7 c_2 }{5} \right) \frac{\varepsilon \, C^{3/2}}{(\alpha^2/ C)^{3/2}} \Lambda_0^2. \label{eq:W3_final} \eeq Using the obtained expressions for $W_2$ and $W_3$ from Eqs.\ (\ref{eq:w2_final}) and (\ref{eq:W3_final}),  we obtain the expression for the velocity derivative skewness as \beq S=\frac{W_3}{W_2^{3/2}}= \frac{\sqrt{10}}{4 \pi^2} \frac{\frac{51 c_1}{392}+\frac{7c_2}{5}}{\left[1+\frac{7}{300 (\alpha^2/C)}\right]^{3/2}}  \frac{1}{(\alpha^2/C)^{3/2}}.\label{eq:skew}\eeq The amplitude ratio $\alpha^2/C=0.050$ is calculated in Sec.\ \ref{sec:RSD}, and $c_1$ and $c_2$ are two constants whose values namely, $c_1=0.553$, and $c_2=-0.166$, are obtained via numerical integrations as shown in the Appendix C and D. Using  these numerical values in Eq.\ (\ref{eq:skew}), we obtain \beq S=-0.647.\eeq This value of velocity derivative skewness is comparable with recent DNS and experimental values. In Table 1, various theoretical, experimental, and numerical predictions for skewness are displayed for comparison.
\section{Evaluation of skewness including the dissipation range}
It is well known that following the inertial range there occurs a dissipation range where dissipative effects due to the viscosity are dominant. It would be interesting to obtain the skewness value taking the dissipative effects into account. However, there exists no standard theory for the dissipation range and the renormalization schemes including the closure approximations have not been able to address the behavior of this range. Such schemes so far have calculated only inertial range quantities where power like spectra exist. There has been a lot of studies in the structure of fine scale turbulence within the inertial range. The corresponding statistical characterization is important in the sense that Kolmogorov 1941 theory of universal range would predict a skewness independent of the Reynolds number. 

Despite the above fact, we expect that a renormalized theory can be  extended to include the dissipation range, an example being the  EDQNM formulation. Since our formulation is based on renormalized quantities, we expect a similar kind of extension to be valid. 

The energy spectrum $E(k)$ including the dissipative effects has been modeled in different ways in the literature. Here we take the model of Pao \cite{Pao_POF_1965}, namely,  \beq E(k) = C \varepsilon^{2/3} k^{-5/3} e^{- \beta (k \eta)^{4/3}}, \label{eq:pao}\eeq where $\beta=2.400$ for $C=1.600$ and $\eta$ is the Kolmogorov dissipation length scale defined as $\eta = (\nu^3_0/\varepsilon)^{1/4}$ \cite{Smith92}. In this section we shall denote the second and third cumulants as $\mathcal{W}_2$ and $\mathcal{W}_3$ that include the contributions due to the dissipation range.
\subsection{Evaluation of $\mathcal{W}_2$}
We generalize the expression for $W^{(1)}_2$ as \beq  \mathcal{W}^{(1)}_2= \frac{S_{d-1} 2\pi^2  C \varepsilon^{2/3}}{[2\pi]^d} \int^{\infty}_0 \ dk \ k^{1/3} e^{- \beta (k\eta)^{4/3}} \int^{\pi}_0 d\theta \cos^2\theta (1-\cos^2\theta) \sin\theta  \label{eq:W2_1_an_wav_dissi} \eeq that incorporates the dissipation range because the energy spectrum due to Pao [Eq.\ (\ref{eq:pao})] has been employed. The ultra-violet limit has been extended to infinity to include all dissipative effects occurring in the small scales of motions. We make the change of variable as $ \beta^{3/4} k\eta=s$ and obtain \beq \mathcal{W}^{(1)}_2= \frac{2 a}{15} \frac{C \varepsilon^{2/3}}{\beta \eta^{4/3}},\label{eq:w21_eta}\eeq where \beq a=\int^{\infty}_0 ds  \ s^{1/3} e^{- s^{4/3}}. \label{eq:a} \eeq We make a similar generalization of the expression for $W^{(2)}_2$ to incorporate Pao's model and obtain \beq \mathcal{W}^{(2)}_2= \frac{7}{2250}  \frac{C^2 \varepsilon^{2/3}}{\alpha^2} \int^{\infty}_0 \ dk \ k^{1/3} e^{- \beta (k\eta)^{4/3}}. \label{eq:W22_eta1} \eeq Using the same change of variables, it can be written as \beq \mathcal{W}_2^{(2)}= \frac{7 \, a}{2250 (\alpha^2/C)} \, \frac{C \varepsilon^{2/3}}{\beta \eta^{4/3}}\label{eq:w22_eta3}\eeq so that \beq \mathcal{W}_2=\mathcal{W}_2^{(1)} +\mathcal{W}_2^{(2)}=\frac{2 \, a}{15} \left[1+\frac{7}{300 \, (\alpha^2/C) }\right]  \frac{C \varepsilon^{2/3}}{\beta \eta^{4/3}}.\label{eq:w2_final2}\eeq 
\subsection{Evaluation of $\mathcal{W}_3$}
Here we generalize the expression for $W^{(1)}_3$ to include the dissipation range and obtain \beq
\mathcal{W}_3^{(1)}= \frac{51 \, \, b_1}{15680 \pi^2 (\alpha^2/C)^{3/2}} \ \frac{\varepsilon \ C^{3/2}}{\beta^{3/2}\eta^{2}} \eeq with $$ b_1= \int^{\infty}_0 d s \int^{\infty}_0 d s' \int^\pi_0 \sin \theta d\theta \int^\pi_0 \sin \theta' d\theta' \int^{2\pi}_0 d \phi \int^{2\pi}_0 d \phi' \  $$ \beq M (\mb s,\mb s',\theta,\theta', \phi, \phi') \ e^{-3 s^{4/3}/4 } e^{-3 s'^{4/3}/4}, \label{eq:Ilambda_dissi} \eeq where $s=\beta^{3/4} k\eta$ and $s'=\beta^{3/4} k'\eta$.

In a similar way as above, we generalize the expression for $W^{(2)}_3$ to include the dissipative effects as  \beq \mathcal{W}_3^{(2)}=- \frac{7 \ b_2}{200 \pi^2} \ \frac{1}{(\alpha^2/C)^{3/2}} \frac{\varepsilon \ C^{3/2}}{\beta^{3/2}\eta^{2}}, \label{eq:w32_3_dissi}\eeq where $$ b_2= \int^{\infty}_0 d s  \int^{\infty}_0 ds' \int^\pi_0 \sin \theta \, d\theta \int^\pi_0 \sin \theta'\, d\theta' \int^{2\pi}_0 d \phi \int^{2\pi}_0 d \phi'  \, R_2(\mb s,\mb s') \,s^{-y} \ s'^{-11/3} $$ \beq T(\mb s,\mb s',\theta_1,\theta_2,\phi_1,\phi_2)  e^{-(3/4)\ s^{4/3}} \ e^{-(3/4) s'^{4/3}}. \label{eq:W23_int_dissi} \eeq The contributions from $\mathcal{W}_3^{(3)}$ and $\mathcal{W}_3^{(4)}$   are irrelevant because they cancel each other out in this case also. 
\subsection{Evaluation of Skewness} 
We obtain the total contribution to $\mathcal{W}_3$ as \beq \mathcal{W}_3=\mathcal{W}_3^{(1)}+\mathcal{W}_3^{(2)} =\frac{1}{40 \pi^2} \left[\frac{51 b_1}{392}+\frac{7 b_2}{5} \right] \frac{1}{(\alpha^2/ C)^{3/2}} \frac{\varepsilon C^{3/2} }{\beta^{3/2}\eta^2}. \label{eq:W3_dissi_final} \eeq The resulting expression for skewness, from Eqs.\ (\ref{eq:w2_final2}) and (\ref{eq:W3_dissi_final}) turns out to be \beq \mathcal{S}=\frac{ \mathcal{W}_3 }{\mathcal{W}^{3/2}_2}=\frac{3\sqrt{15}}{16 \sqrt{2} \,a^{3/2}\pi^2} \frac{\left(\frac{51 \ b_1}{392}+\frac{7 \ b_2}{5}\right)}{\left[1+\frac{7}{300 (\alpha^2/C)}\right]^{3/2}} \frac{1}{(\alpha^2/C)^{3/2}}.\label{eq:skew_dissi}\eeq We observe that the skewness value $\mathcal{S}$ depends on the parameters $a$, $b_1$, $b_2$ and $\alpha^2/C$. Evaluating numerically the constants  $a$, $b_1$   and $b_2$ from the integrals given by Eqs.\ (\ref{eq:a}), (\ref{eq:Ilambda_dissi}) and  (\ref{eq:W23_int_dissi}), we obtain $a=0.750$, $b_1=0.928$ and $b_2=-0.207$. Using them in the above expressions for $\mathcal{S}$ we obtain \beq \mathcal{S}=-0.682. \eeq It is interesting to observe that the skewness value does not change drastically from the inertial range value when the dissipation range is included. In fact the magnitude of skewness acquires a slightly higher value than the inertial range value.  
   
\section{Discussion and Conclusion}
In this paper, we obtained renormalized expressions for the second- and third-order cumulants of velocity derivative by applying a renormalized perturbative scheme on the NS equation for an incompressible isotropic turbulent velocity field. This scheme of calculation finds out the renormalized quantities directly from various loop diagrams for the second and third order cumulants of velocity derivative. This type of scheme has previously been used for the calculation of statistical cumulants in KPZ \cite{Singha-Nandy14_PRE.90.062402,Singha-Nandy15_jstatmech,Singha-Nandy_KPZ2D_JSTAT} and VLDS \cite{Singha-Nandy16_jstatmech} surface growth dynamics. Employing diagrammatic approach, we have seen that there are two contributing Feynman diagrams [Fig.\ \ref{fig:w2}] at one-loop order for the second cumulant $W_2$. We evaluated the amputated part of the loop-diagram appearing in Fig.\ \ref{fig:w2}(b) as given by Eq.\ (\ref{eq:w2_final}). In total, there are five Feynman diagrams for the third cumulant $W_3$ as shown in Fig.\ \ref{fig:w3}. Calculating each of the diagrams, we have seen that one diagram, namely, Fig.\ \ref{fig:w3}(c), gives vanishing contribution. Further, Figs.\ \ref{fig:w3}(d) and \ref{fig:w3}(e), corresponding to $W_3^{(3)}$ and $W_3^{(4)}$, yield logarithmic contributions with opposite signs and they cancel each other out. The remaining two diagrams, namely, Figs.\ \ref{fig:w3}(a) and \ref{fig:w3}(b), finally leads to a negative value of $W_3$ as given by Eq.\ (\ref{eq:W3_final}). This result, combined with the result for $W_2$ given by Eq.\ (\ref{eq:w2_final}), yields the expression for velocity derivative skewness  given by Eq.\ (\ref{eq:skew}) when only the inertial range, namely, the energy spectrum given by  Eq.\ (\ref{Eq:energy_Kol}), is employed and the dissipation range is neglected. We see that the resulting expression for skewness depends on three constants $c_1$, $c_2$ and $\alpha^2/C$. We numerically evaluated the integral expressions determining $c_1$ and $c_2$ yielding $c_1=0.553$ and $c_2=-0.166$. The value of amplitude ratio $\alpha^2/C$ is determined employing the same procedure, yielding $\alpha^2/C=0.050$. Using these values, we obtained the skewness value  as $S=-0.647$. 
\begin{table}[h]
\caption{Various experimental, numerical, and theoretical estimates for the velocity derivative skewness in three-dimensional homogeneous isotropic turbulence. The value shown against Ref.\ \cite{Gylfason04} is evaluated from their predicted scaling law. The number of significant digits displayed are according to the availability from the sources.}\footnotesize
\begin{center} 
\begin{tabular}{l l l l } 
\hline\hline 
Method & $R_\lambda$ & Ref.  & Skewness \\
\hline
Rotating disk & 200 to 700 & \cite{Belin97}  & $-0.5$ \\

Hot-wire& 149 to 729 & \cite{Gylfason04} & $-0.518$ to $-0.597$\\ 
 
Hot-wires & 11 to 47 & \cite{Burattini2008} & $-0.5$ ($\pm5\%$) \\
 
Numerical & 20 to 45 & \cite{Orszag72} & $-0.47$ \\
 
Numerical& 28.9 to 82.9 & \cite{Kerr85} &$-0.505\pm 0.005$ \\
 
DNS & $150$ & \cite{Vincent91} & $-0.5$ \\ 
 
DNS & 21 to 195 & \cite{Wang96} & $-0.5$ \\
 
DNS & 38 to 460 & \cite{Gotoh02} & $-0.53$ \\
 
DNS & $680$ & \cite{Ishihara07} & $-0.648\pm 0.003$ \\
 
EDQNM & asymptotic & \cite{Andre77}&  $-0.495$ \\
 
EDQNM  & asymptotic & \cite{Lesieur00} & $-0.547$ \\

RG & asymptotic & \cite{YO_86} &  $-0.4878$ \\ 
 
RG & asymptotic & \cite{Smith92} &$-0.59$ \\
 
MSCE & asymptotic & \cite{Tatsumi78} & $-0.65$ \\ 
 
MLRA & asymptotic & \cite{Kaneda93} &$-0.66$ \\
 
LDIA & asymptotic & \cite{Kida97} & $-0.66$  \\

Present scheme (excluding dissipation) & asymptotic & Eq.\ (\ref{eq:skew}) &$-0.647$ \\

Present scheme (with dissipation) & asymptotic & Eq.\ (\ref{eq:skew_dissi}) &$-0.682$  \\\hline
\hline 
\end{tabular} \end{center} \label{tab:skew}
\end{table}

It is however important to take the dissipation range into account in order to calculate the integrals for the second and third cumulants, $W_2$ and $W_3$. To do so, we employed Pao's model with the energy spectrum given by Eq.\ (\ref{eq:pao}) that joins the inertial range smoothly with the dissipation range. We repeated the calculations and obtained the expression for the corresponding skewness value $\mathcal{S}$ as given by Eq.\ (\ref{eq:skew_dissi}). This expression for $\mathcal{S}$ depends on the constants $a$ , $b_1$, $b_2$ and $\alpha^2/C$. We evaluated the integral expressions determining $a$, $b_1$ and $b_2$ yielding $a=0.750$, $b_1=0.928$ and $b_2=-0.207$. The skewness value turns out to be $\mathcal{S}=-0.682$. We observe that this value is somewhat close to the inertial range skewness value giving us the impression that inclusion or exclusion of the dissipation range does not affect the skewness value drastically.

As shown in Table 1, our present estimate ($S=-0.647$) is relatively higher in magnitude than previous RG estimates of Yakhot and Orszag ($S=-0.4878$) and Smith and Reynolds ($S=-0.59$). This may be attributed to the fact that our present  scheme finds a relation  between the renormalized Feynman diagrams involving the renormalized viscosity (instead of bare viscosity). Accordingly, the velocity correlation appearing in the Feynman diagrams for the cumulants are expressed in terms of the renormalized quantities given by Eq.\ (\ref{vel-corre}). This procedure further allowed us to estimate the amplitude ratio $\alpha^2/C$ which turned out to be very low ($\alpha^2/C=0.05$) compared to that of the Yakhot and Orszag  ($\alpha^2/C\approx 0.15$). Since this amplitude ratio appears in the denominator of the final expression for velocity derivative skewness [Eq.\ (\ref{eq:skew})], we obtained a comparatively higher value for the skewness. Our present result for velocity derivative skewness also differ from that of EDQNM estimates \cite{Andre77,Lesieur00}. In EDQNM formalism, the velocity derivative skewness is completely determined by the second-order moments whereas, our present scheme takes into account all the relevant Feynman diagrams for second and third cumulants.

Our present theoretical skewness values are in the vicinity of the other theoretical estimates coming from MSCE \cite{Tatsumi78}, and the closure theories, namely, LRA \cite{Kaneda93} and LDIA \cite{Kida97}. Our present estimates is also comparable to the recent estimates from a high resolution DNS ($4096^3$ grid points) performed by  Ishihara {\it et al.}\ \cite{Ishihara07}, giving $S=-0.648 \pm 0.003$  for $R_\lambda=680$. Their DNS  suggested an empirical relation $S\sim -(0.32 \mp 0.02) \ R_{\lambda}^{0.11\pm 0.01}$, obtained via a least square fit of the DNS data in the range $200 \leqslant R_\lambda \leqslant 680$. In fact, their DNS data for $S$ up to $R_{\lambda}=1130$ was consistent with Gylfason's \cite{Gylfason04} scaling relation. The scaling relations between $S$ and $R_\lambda$ support that the skewness is a mildly growing function of Reynolds number. We would like to note that we have calculated the second- and third-order velocity derivatives, and hence the skewness, assuming the Kolmogorov phenomenology and Pao's modification for inclusion of the dissipation range to be valid. These phenomenological considerations are in fact valid for infinitely large Reynolds numbers, suggesting  that our calculations for skewness correspond to infinite Reynolds number. Thus it is difficult to compare them with the experimental and numerical results that are usually obtained for finite (although high) Reynolds numbers. As discussed above, experimental and numerical estimates for skewness have been expressed in the form $S=- \sigma R^{\delta}_{\lambda}$ (with $\sigma$ and $\delta$ positive). This empirical relation is usually valid for a (finite) range of high $R_{\lambda}$ values beyond which its validity is unknown. However, this scaling law, if assume to be extended to Reynolds numbers higher than those considered in the experiments and numerical simulations, the skewness would grow indefinitely (although slowly) with increasing Reynolds numbers. This situation appears to be quite unlikely as the skewness can not be infinitely large \cite{Lesieur}. Our theoretical estimates, on the other hand, indicate that the skewness is a finite quantity in the limit of infinite Reynolds number. In fact, Tatsumi {\it et al.}\ \cite{Tatsumi78} showed through the MSCE method that the skewness value saturates to a constant value as the Reynolds number increases boundlessly. Our calculated skewness values $S=-0.647$ and $\mathcal{S}=-0.682$ compares well with Tatsumi's asymptotic value $-0.65$ for  infinitely large Reynolds number. It can be guessed that beyond the scaling regime ($S\sim R^{\delta}_{\lambda}$) observed in experiments and numerical simulations, the skewness ought  to saturate to a constant value. 
\section*{Appendix A. Calculation of $W^{(1)}_2$}
We perform the integration over frequency in Eq.\ (\ref{eq:W21-Fourier1}) and obtain \beq W_2^{(1)}= \frac{S_{d-1}D_0}{[2\pi]^d} \int^{\Lambda_0}_0 dk \frac{k^{d-y+1}}{\zeta(k)} \int^\pi_0 \sin^{(d-2)}\theta \cos^2\theta (1-\cos^2\theta) d\theta, \label{eq:W2_1_an_wav} \eeq where $\zeta(k)$ is related to the renormalized viscosity $\nu(k)$ as $\zeta(k)=k^2 \nu(k)$. Carrying out the angular and wave vector integrations and substituting $D_0=2\pi^2\alpha C \varepsilon$, we arrive at Eq.\ (\ref{eq:w21_3}) in three dimensions. 
\section*{Appendix B. Calculation of $W^{(2)}_2$}
Equation\ (\ref{eq:L2}) yields \beq K^{(2)<}_{mnsl}(r)=\frac{S_d}{[2 \pi]^d} \frac{(d^2-2)}{2d(d+2)}  \left(\frac{\lambda^2_0 D^2_0}{\nu_0^3}\right)  \left[\frac{\Lambda^{d-2y-6}_0-(\Lambda_0 e^{-r})^{d-2y-6}}{d-2y-6}\right]\delta_{ml} \delta_{ns},\eeq which leads to the differential equation \beq \frac{dK^{(2)}_{mnsl}(r)}{dr}= \frac{S_d}{[2\pi]^d}  \frac{(d^2-2) \lambda^2_0  D^2_0}{2d(d+2)\nu^3(r)} (\Lambda_0 e^{-r})^{d-2y-6} \delta_{ml} \delta_{ns}.\eeq Integrating with respect to $r$, we obtain \beq K^{(2)}_{mnsl}(r)= \frac{S_d}{[2\pi]^d} \frac{(d^2-2)}{2d(d+2)(y+2)} \left(\frac{\lambda^2_0 D^2_0}{\nu^3(r)}\right)(\Lambda_0e^{-r})^{d-2y-6} \delta_{ml} \delta_{ns}. \eeq Substituting $y=d$ in three dimensions, we obtain 
\beq K^{(2)}_{mnsl}(r)=\frac{7\lambda^2_0 D^2_0}{300\pi^2\nu^3(r)} (\Lambda_0 e^{-r})^{-9} \delta_{ml} \ \delta_{ns}.  \label {eq:L2_f}\eeq

Now we construct $k$ and $\omega$ dependences of $K^{(2)}_{mnsl}(k, \omega)$ by identifying  $\Lambda_0 e^{-r}$ with 
$k$ and a dimensionless scaling function is employed to obtain $(\Lambda_0 e^{-r})^{-5}$ as \beq (\Lambda_0 e^{-r})^{-5}=k^{-1} \nu^{2}(k) |G(k,\omega)|^2 \eeq which has the desired limit $k^{-5}$ in the limit of $\omega \rightarrow 0$. Thus, Eq.\ (\ref{eq:L2_f}) can be expressed as \beq K^{(2)}_{mnsl}(k,\omega)=\frac{7 \pi^2 \varepsilon C^2}{75 \alpha }k^{-1} \nu^{2}(k) |G(k,\omega)|^2\delta_{ml} \delta_{ns}.\label{eq:L2_k}\eeq Substituting Eq.\ (\ref{eq:L2_k}) in Eq.\ (\ref{eq:W22}) and performing frequency and angular integrations, we obtain \beq W^{(2)}_2= \frac{7}{2250} \frac{C^2 \varepsilon^{2/3} }{\alpha^2} \int^{\Lambda_0}_0 q^{1/3} dq. \label{eq:W2_2wav}\eeq in three dimensions, leading to the result of Eq.\ (\ref{eq:w22_3}).
\section*{Appendix C. Calculation of $W_3^{(1)}$} \label{sec:w3_1}
Using the expression for noise correlation as given by Eq. (\ref{eq:force}), we evaluate the integral appearing in Eq.\ (\ref{eq:L0}) with the assumption that the internal wave-vector $\mb q$ is much greater in magnitude than external wave-vectors. Consequently we obtain \beq L^{(1)}_{ijlmns}= i \lambda^3_0 (2D_0)^3 \int \frac{d^d q}{[2\pi]^d}\int\frac{d\Omega}{[2\pi]} P_{mj}(\mb q) P_{ns}(\mb q) P_{il}(\mb q) |G_0(\mb q,\Omega)|^6 |\mb q|^{-3y},\eeq which, upon performing frequency convolution, yields \beq L^{(1)}_{ijlmns}= \frac{3}{2}\frac{i \lambda^3_0 D_0^3}{\nu^5_0} \int \frac{d^d q}{[2 \pi]^d} \ P_{mj}(\mb q) P_{ns}(\mb q) P_{il}(\mb q) \ |\mb q|^{-3y-10}.\eeq Eliminating the high wave number band $\Lambda_0 e^{-r} \leqslant q \leqslant \Lambda_0$, we obtain \beq L^{(1)<}_{ijlmns}(r)= \frac{S_d}{[2\pi]^d} \frac{3i \lambda^3_0 D_0^3}{2\nu^5_0} F^{(1)}_{ijlmns}(d) \int^{\Lambda_0}_{\Lambda_0 e^{-r}} q^{d-3y-11} dq, \eeq  where we define \bea F^{(1)}_{ijlmns}(d)&=&  [f_1(d)\delta_{mj}\delta_{ns}\delta_{il}+f_2(d)(\delta_{ms}\delta_{nj} \delta_{il}+\delta_{mj}\delta_{is}\delta_{nl} +\delta_{im}\delta_{ns} \delta_{jl}) \nonumber \\ &-&f_3(d)(\delta_{ml}\delta_{in}\delta_{js}+ \delta_{ms}\delta_{in}\delta_{jl} +\delta_{im}\delta_{nl}\delta_{js}+\delta_{ml} \delta_{is}\delta_{nj}) ] \eea with \[ f_1(d)=1-\frac{3}{d}+\frac{3}{d(d+2)}-\frac{1}{d(d+2)(d+4)}, \hspace{0.5cm} f_2(d)=\frac{1}{d(d+2)}-\frac{1}{d(d+2)(d+4)},\] \beq f_3(d)=\frac{1}{d(d+2)(d+4)}. \eeq

We consider the iterative nature of the shell elimination scheme in thin shells in the wave-vector space and obtain the flow of $L^{(1)}_{ijlmns}(r)$ in the form of a differential equation \beq \frac{dL^{(1)}_{ijlmns}}{dr}=\frac{S_d}{[2\pi]^d}  \frac{3i \lambda^3_0D^3_0}{2\nu^5(r)} \Lambda^{d-3y-10}(r) F^{(1)}_{ijlmns}(d). \eeq Solving this equation in the asymptotic limit of large $r$, we obtain for $y=d$ \beq L^{(1)}_{ijlmns}(r)=\frac{S_d}{[2\pi]^d}  \frac{3i \lambda_0 D^2_0}{2(2d+10/3)} \left(\frac{\lambda^2_0  D_0}{\alpha^3\ \varepsilon}\right) \frac{\Lambda^{-2d-6}(r)}{\nu^2(r)} F^{(1)}_{ijlmns}(d). \eeq

To find the wave-vector and frequency dependence, we identify $\Lambda_0 e^{-r}$ with $k$ and a dimensionless scaling function is employed to obtain $(\Lambda_0 e^{-r})^{(-d-5/3)}$ as \beq k^{(7-3d)/3} \nu^2(k)  |G(\hat{k})|^2, \eeq yielding the renormalized amputated loop diagram in Fig.(2a) as \beq L^{(1)}_{ijlmns}(\mb k,\omega,\mb k',\omega')=\frac{S_d}{[2\pi]^d}  \frac{3 i (2 \pi^2 \alpha C \varepsilon)^3 }{2 \alpha \ \varepsilon^{1/3} (2d+10/3) } F^{(1)}_{ijlmns}(d) k^{-(d+1/3)}  k'^{-(d+1/3)} |G(\hat{k})|^2 |G(\hat{k'})|^2. \label{eq:Lk}\eeq Substituting Eq.\ (\ref{eq:Lk}) in Eq.\ (\ref{eq:w3a}) and carrying out the frequency integrations, we obtain \[W_3^{(1)}=- \frac{S_d}{[2\pi]^d} \frac{3 (2 \pi^2 \alpha C \varepsilon)^3 }{2 \alpha \ \varepsilon^{1/3} (2d+10/3) } \int \frac{d^d k}{[2\pi]^d} \int \frac{d^d k'}{[2\pi]^d}  k_3 k'_3 (k_3+k'_3) \]\beq [P_{3mn}(\mb k) P_{3ij}(\mb k')P_{3sl}(-\mb k-\mb k')] \ k^{-(d+1/3)} \ k'^{-(d+1/3)} R(\mb k, \mb k'), \label{pre-mom-int} \eeq where \beq R(\mb k, \mb k')=\frac{N(\mb k,\mb k')}{D(\mb k,\mb k')} \eeq with \beq N(\mb k,\mb k')= 3[\zeta^2(k)+\zeta^2(k')]+4 \zeta(|\mb k+\mb k'|)[\zeta(k)+\zeta(k')]+\zeta^2(|\mb k+\mb k'|)+14 \zeta(k) \zeta(k'), \eeq and \beq D(\mb k,\mb k')=16 \ \zeta^2(k) \ \zeta^2(k') \left[\zeta(k)+\zeta(k')+\zeta(|\mb k+\mb k'|)\right]^3. \eeq For $d=3$, tensorial contraction leads to \beq F^{(1)}_{ijlmns}(d) [P_{3mn}(\mb k) P_{3ij}(\mb k')P_{3sl}(\mb k+\mb k')] =\frac{34}{105}[P_{3js}(\mb k) P_{3jl}(\mb k') P_{3sl}(\mb k+\mb k')], \eeq yielding \beq W_3^{(1)}= \frac{S_d}{[2\pi]^d} \, \frac{17\lambda_0(2\pi^2 \alpha C \varepsilon)}{35[2\pi]^{2d}(2d+10/3)} \left(\frac{\lambda_0 2\pi^2 C }{\alpha^2\ } \right)^2\ I_1(\Lambda_0),\label{eq:WWW}\eeq with \beq I_1(\Lambda_0)= \int^{\Lambda_0}_0 dk \int^{\Lambda_0}_0 dk' \int^\pi_0 \sin \theta d\theta \int^\pi_0 \sin \theta' d\theta' \int^{2\pi}_0 d \phi  \int^{2\pi}_0 d \phi' \  M (\mb k,\mb k',\theta,\theta', \phi, \phi'), \label{eq:Ilambda}\eeq where \beq M (\mb k,\mb k',\theta,\theta', \phi, \phi')=-R'(\mb k,\mb k')  \ k^{-4/3} \ k'^{-4/3} J(\mb k,\mb k',\theta,\theta',\phi,\phi'), \eeq with \beq J(\mb k,\mb k',\theta,\theta',\phi,\phi')= k_{3} k'_{3} (-k_{3}-k'_{3})[P_{3js}(\mb k) P_{3jl}(\mb k')P_{3sl}(-\mb k-\mb k')]. \label{eq:J} \eeq and \beq R'(\mb k, \mb k')= (\alpha \varepsilon^{1/3})^5 R(\mb k, \mb k').\eeq The expression given by Eq.\ (\ref{eq:J}) can be simplified by using the spherical polar coordinate where $k_{13}=k_1 \, x=k_1 \cos\theta_1$, $k_{23}=k_2 \, y=k_2 \cos\theta_2$ and $\mb k_1 \cdot \mb k_2= k_1 \, k_2 \, z=k_1 \, k_2 
\left[\sin\theta_1 \sin\theta_2\cos(\phi_1-\phi_2)+\cos\theta_1\cos\theta_2 \right]$. Thus, we obtain
\[ J(k_1,k_2,\theta_1,\theta_2,\phi_1,\phi_2)= 2 |k_1|^3 |k_2|^3 \left[\rho_1(x,y,z)+ \rho_2(x,y,z) \frac{(k_1 x+k_2 y)^2}{k_1^2+k_2^2+2 k_1 k_2 z}\right ] \] \[+2 |k_1|^2 |k_2|^4 \left[\sigma_1(x,y,z)+\sigma_2(x,y,z) \frac{(k_1 x+k_2 y)^2}{k_1^2+k_2^2+2 k_1 k_2 z}\right] \]
\beq+2 |k_1|^4 |k_2|^2 \left[\tau_1(x,y,z)+\tau_2(x,y,z) \frac{(k_1 x+k_2 y)^2}{k_1^2+k_2^2+2 k_1 k_2 z}\right], \eeq
where  
\[\rho_1(x,y,z)=x(y^3-y^5-y^3 z^2)+2x^2(y^2 z-y^4 z)+x^3(4 y^3 z^2+y-y z^2)-2x^4 y^2 z-x^5 y, \]
\[\rho_2(x,y,z)=x y \left[x^2(1+4 z^2)+y^2(1+4 z^2)-z^2-1-4 x y z (1+z^2)\right],\]
\[\sigma_1(x,y,z)=x(y^3 z-2y^5 z)+ x^2(y^2-y^2 z^2+2 y^4 z^2)-x^4 y^2,\]
\[\sigma_2(x,y,z)=xy \left[2x^2 z+z(2 y^2-1)-x(y+2yz^2) \right],\]
\[\tau_1(x,y,z)=x^2(y^2+y^2 z^2-y^4)+ x^3 y z+2x^4 y^2 z^2,\]
\[\tau_2(x,y,z)=xy \left[2x^2 z+z(2 y^2-1)-x(y+2yz^2) \right].\]
Scaling the wave numbers with respect to $\Lambda_0$ and performing the integration numerically in Eq.\ (\ref{eq:Ilambda}) for large ultra-violet limits, we obtain \beq c_1=\lim_{\Lambda_0 \rightarrow \infty} [I_1(\Lambda_0)/\Lambda_0^2]= 0.553. \label{eq:c_1} \eeq We thus obtain the result of Eq.\ (\ref{eq:final_W31}).

\section*{Appendix D. Calculation of $W_3^{(2)}$}
Performing frequency integration on Eq.\ (\ref{eq:calL}), we obtain \beq L^{(2)<}_{mnab}(r)= \frac{S_d}{[2 \pi]^d} \frac{1}{4 \nu^3_0}  \ F^{(2)}_{mnab}(d) \ \left[\frac{\Lambda^{d-12}_0-(\Lambda_0 e^{-r})^{(d-12)}}{d-12}\right], \eeq where \beq F^{(2)}_{mnab}(d)=\left[\delta_{ma}\delta_{nb}-\frac{2}{d} \delta_{ma} \delta_{nb}+\frac{\delta_{ma} \delta_{nb}+\delta_{mb} \delta_{na}+\delta_{mn} \delta_{ab}}{d(d+2)}\right]. \eeq Using a similar procedure as in Sec.\ II of the appendix, we obtain renormalized amputated loop as \beq L^{(2)}_{mnab}(r)=\frac{S_d}{[2\pi]^d}  \frac{1}{20 \alpha^3 \varepsilon}  F^{(2)}_{mnab}(d) \Lambda^{-5}(r). \eeq For wave vector and frequency dependences, we identify $(\Lambda_0 e^{-r})^{-5}$ as \beq p^{-1} \, \nu(p)^2 \, |G(p,\Omega)|^2,\label{eq:DS2}\eeq so that we write  \beq L^{(2)}_{mnab}(\mb p,\Omega)=\frac{S_d}{[2 \pi]^d}   \frac{F^{(2)}_{mnab}(d)}{20 \ \alpha \varepsilon^{1/3}} \,  p^{-11/3}|G(p,\Omega)|^2. \eeq Substituting this expression in Eq.\ (\ref{eq:w3b}), we obtain $$W_3^{(2)}=  \frac{6}{5} \frac{S_d}{[2\pi]^d}  \frac{(2\pi^2 \alpha C \varepsilon)^3 F^{(2)}_{mnab}(d)}{\ \alpha \varepsilon^{1/3}} \int \frac{d^{d+1}\hat{k}}{[2\pi]^{d+1}} \int \frac{d^{d+1} \hat{p}}{[2\pi]^{d+1}} k_3 p_3 (k_3+p_3)P_{3mn}(\mb p)P_{3ij}(\mb k+\mb p)$$\beq  P_{iab}(-\mb p) P_{3j}(\mb k) |k|^{-y}  \ |p|^{-11/3} |G(\hat p)|^2 |G(\hat k)|^2 G(-\hat k-\hat p)  |G(\hat p)|^2. \label{eq:w3b1}\eeq

We carry out the frequency integrations in Eq. (\ref{eq:w3b1}) and obtain \beq W_3^{(2)}= \frac{56}{25} \pi^4 \alpha^2 C^3  \varepsilon^{8/3} \int\frac{d^3k}{[2\pi]^3}\int\frac{d^3p}{[2\pi]^3} R_2(\mb k,\mb p) |k|^{-y}|p|^{-11/3} \, T(\mb k,\mb p,\theta_1,\theta_2,\phi_1,\phi_2), \label{bef-int-sim-W23} \eeq in $d=3$ where \beq R_2(\mb k,\mb p)=\frac{\zeta(k)+2 \zeta(p)+\zeta(|\mb k+\mb p|)}{8 \, \zeta(k) \zeta^3(p) \{\zeta(k)+ \zeta(p)+\zeta(|\mb k+\mb p|)\}^2}, \eeq coming from frequency integrations and \beq T(\mb k,\mb p,\theta,\theta',\phi,\phi')=2 (p_3 p_i-\delta_{3i}p^2) P_{3ij}(\mb k+\mb p)P_{3j}(\mb k). \label{eq:T} \eeq The expression given by Eq.\ (\ref{eq:T}) can be expressed using spherical polar coordinate as  \[ T(k,p,\theta_1,\theta_2,\phi_1,\phi_2) =2kp^5 \rho_3(x,y,z)+ 2 k^2 p^4 \sigma_3(x,y,z)+ 2 k^3 p^3 \tau_3(x,y,z) \] \beq-{4 k^2 p^4 \frac{(p x+k y)^2}{k^2+p^2+2 k p z} \xi(x,y,z)},\eeq
where 
\[\rho_3(x,y,z)=(x^2 y^2 z-x^3 y-x^4 y^2 z+x^5 y),\]
\[\sigma_3(x,y,z)=(x y^3 z+x^2 y^4-2x^2 y^2+x^3 y z-x^3 y^3 z+x^4 y^2),\]
\[\tau_3(x,y,z)= (x^2 y^2 z-x^2 y^4 z -x y^3+x y^5),\]
\[\xi(x,y,z) =x y\left[x^2 z+y^2 z-xy(1+z^2)\right].\]
Equation \ (\ref{bef-int-sim-W23}) is expressed as \beq W_3^{(2)}= \frac{7}{200 \pi^2} \frac{\varepsilon \  C^{3/2}}{(\alpha^2/C)^{3/2}} \, I_2(\Lambda_0),\eeq where $$ I_2(\Lambda_0)=\int^{\Lambda_0}_0 dk  \int^{\Lambda_0}_0 dp \int^\pi_0 \sin \theta \, d\theta \int^\pi_0 \sin \theta'\,d\theta' \int^{2\pi}_0 d \phi  \int^{2\pi}_0 d \phi' \, R'_2(\mb k,\mb p) \, $$ \beq k^{-y} p^{-11/3} \,  T(\mb k,\mb p,\theta,\theta',\phi,\phi') \label{eq:W3_2_int}\eeq  with \beq R'_2(\mb k,\mb p)=\alpha^5 \varepsilon^{5/3} R_2(\mb k,\mb p).\eeq

We carry out the integrations in Eq. (\ref{eq:W3_2_int}) numerically for large ultra-violet limits and obtain \beq c_2=\lim_{\Lambda_0 \rightarrow \infty} [I_2(\Lambda_0)/\Lambda_0^2]= -0.166. \label{eq:c_2} \eeq Thus, we obtain the result of Eq.\ (\ref{eq:final_W32}). 

\section*{Appendix E. Calculation of $W^{(3)}_3$}
Assuming the internal wave number $q$ to be much greater than external wave numbers $k$ and $k'$, we obtain from Eq.\ (\ref{eq:L3k'})\beq L^{(3)}_{ibnj}(\mb 0, 0)=2 D_0 \int \frac{d^{d+1} \hat{q}}{[2\pi]^{d+1}} [q_b q_n \delta_{ij}-\frac{q_i q_b q_n q_j}{q^2}] G^2(\hat{q}) |G(\hat{q})|^2 |q|^{-y}. \eeq Carrying out the angular and frequency integrations, we have \beq L^{(3)<}_{ibnj}(r)=- \frac{S_d}{[2\pi]^d} \frac{D_0 (\delta_{ib} \delta_{nj}+\delta_{in} \delta_{bj})}{4 \nu^3_0 d(d+2)} \int^{\Lambda_0}_{\Lambda_0 e^{-r}} dq \, q^{-5}.\eeq Now, following the same procedure as in Sec.\ \ref{sec:w3_1}, we obtain  \beq L^{(3)<}_{ibnj}(r)=- \frac{S_d}{[2\pi]^d} \frac{D_0 (\delta_{ib} \delta_{nj}+\delta_{in} \delta_{bj})}{4\alpha^3 \varepsilon d(d+2)} r. \label{eq:L3} \eeq Considering $\Lambda_0 e^{-r}=k'$, so that $r=\ln (\frac{\Lambda_0}{k'})$, Eq.\ (\ref{eq:L3}) yields \beq L^{(3)}_{ibnj}=-\frac{S_d}{[2\pi]^d} \frac{2\pi^2 C (\delta_{ib} \delta_{nj}+\delta_{in} \delta_{bj})}{4 \alpha^2 \ d(d+2)} \ln \left(\frac{\Lambda_0}{k'}\right).\label{eq:L33-final} \eeq Thus, substituting Eq.\ (\ref{eq:L33-final}) in Eq.\ (\ref{eq:W33}), we obtain the logarithmic correction given by Eq. \ (\ref{eq:w3_3F}).
\section*{Appendix F. Calculation of $W^{(4)}_3$}
Considering the external wave numbers $k$ and $k'$ to be much smaller compared to the internal wave number $Q$, we obtain from Eq.\ (\ref{eq:L4kk'}) This yields \beq L^{(4)}_{ibnj}=2D_0 \int \frac{d^{d+1} \hat{Q}}{[2\pi]^d} G(\hat{Q}) G(-\hat{Q}) |G(\hat{Q})|^2 P_{jmn}(-\mb Q) P_{iab}(\mb Q) P_{ma}(\mb Q).\eeq Performing the angular and frequency integrations in the above expression, we obtain \beq L^{(4)<}_{ibnj}(r)= \frac{S_d}{[2\pi]^d} \frac{D_0 (\delta_{ib} \delta_{nj}+\delta_{in} \delta_{bj})}{2 d(d+2) \nu^3_0} \int^{\Lambda_0}_{\Lambda_0e^{-r}} dQ |Q|^{-5}. \eeq Following the similar procedure as in Sec.\ \ref{sec:w3_1}, we obtain \beq L^{(4)<}_{ibnj}(r)=\frac{S_d}{[2\pi]^d}  \frac{D_0 (\delta_{in} \delta_{bj})}{d(d+2) \alpha^3 \ \varepsilon} r. \eeq This yields \beq L^{(4)}_{ibnj}= \frac{S_d}{[2\pi]^d}  \frac{2 \pi^2 C }{d(d+2) \alpha^2 \ } \ln\left(\Lambda_0/k'\right) \delta_{in} \delta_{bj}. \eeq Substituting the expression for $L^{(4)}_{ibnj}(k')$ in Eq.\ (\ref{eq:W34}), we obtain Eq.\ (\ref{eq:w3_4F}).

\subsection*{Acknowledgements}
T.S. is thankful to the Ministry of Human Resource Development (MHRD), Government of India, for financial support through a scholarship. M.K.N. is indebted to the Indian Institute of Technology Delhi, and particularly to  Prof. Ravishankar and Prof. Senthilkumaran, Department of Physics, for hospitality and for extending various facilities at I.I.T. Delhi.

\end{document}